\documentstyle[12pt]{article}

\def\dfrac{\displaystyle\frac}
\def\i{\imath}
\def\d{\partial}
\def\noi{\noindent}
\def\pp{p_\parallel}
\def\ve{\varepsilon}
\def\bs{\bigskip}

\newcommand{\Eq}[1]{Eq.(\ref{#1})}

\newcommand{\refc}[1]{Ref.~\cite{#1}}
\newcommand{\refs}[1]{Refs.~\cite{#1}}

\newcommand{\bea}{\begin{eqnarray}}
\newcommand{\eea}{\end{eqnarray}}
\newcommand{\be}{\begin{equation}}
\newcommand{\ee}{\end{equation}}
\newcommand{\bc}{\begin{center}}
\newcommand{\ec}{\end{center}}
\newcommand{\ba}{\begin{array}}
\newcommand{\ea}{\end{array}}

\newcommand{\cL}{{\cal L}}

\newcommand{\annp}[3]{{\it  Ann. Phys. (N.Y.) }{{\bf #1} {(#2)} {#3}}}

\newcommand{\fort}[3]{{\it Fortsch. Phys. }{ {\bf #1} {(#2)} {#3}}}

\newcommand{\np}[3]{{\it  Nucl. Phys. }{{\bf #1} {(#2)} {#3}}}
\newcommand{\pr}[3]{{\it Phys. Rev.}{{ \bf #1} {(#2)} {#3}}}

\newcommand{\prl}[3]{ {\it Phys. Rev. Lett.}{{ \bf #1} {(#2)} {#3}}}
\newcommand{\pl}[3]{{\it  Phys. Lett. }{{\bf #1} {(#2)} {#3}}}

\newcommand{\rmp}[3]{{\it  Rev. Mod. Phys.} {{ \bf #1} {(#2)} {#3}}}

\newcommand{\sovphjetp}[3]{{\it Sov. Phys. JETP }{{\bf #1} {(#2)} {#3}}}

\normalsize

\begin{document}

\large
\thispagestyle{empty}
\begin{flushright}                              FIAN/TD/95-11\\
                                                hep-ph/9507404\\
                                                June 1995

\vspace{2cm}
\end{flushright}
\bc
\normalsize
{\LARGE\bf Effective Lagrangian of Thermal QED  with External Magnetic
Field and
the Static Limit of the Polarization Operator}

\vspace{3ex}

{\Large Vadim Zeitlin$^{\dagger}$}

{\large Department of Theoretical Physics, P.~N.~Lebedev Physical
Institute,

  Leninsky prospect 53, 117924 Moscow, Russia}\vspace{5ex}

\ec

\centerline{{\Large\bf Abstract}}

\normalsize

\begin{quote}
A low-temperature expansion of QED one-loop effective Lagrangian valid for
a  wide range of parameters is presented
in a form of finite sums of elementary functions.
Starting from the effective action components of the one-loop polarization
operator responsible for Hall conductivity and Debye screening are obtained.
It is shown that in  a strong background magnetic field the Debye
radius depends on spatial direction.
\end{quote}

\vfill
\noindent
$^\dagger$ E-mail address: zeitlin@lpi.ac.ru

\newpage


\setcounter{page}{1}

\section{Introduction}

 In the vicinity of some stellar objects the magnetic field strength and
the fermion density may be very high \cite{Shklovsky,ShapiroT83} and the
corresponding quantum corrections become important \cite{SUsov,Shabad88}.
Many essential properties of the electron-positron plasma with a constant and
uniform magnetic field may be obtained in the framework of
(3+1)-dimensional
quantum electrodynamics (QED) with chemical potential $\mu$, temperature $T$
and constant and uniform magnetic field $B$.  This theory have been
intensively studied in the last 25 years [5-14] using different approaches and
important results were obtained. At the same time, the
resulting expressions are
usually very cumbersome [9-12] (see discussion in \refc{Shabad88}) and
it is quite difficult even  to match
specific calculations made in different approaches.  For
that reason an approach providing meaningful results in a simple and
easily reproducible form would be useful.

The expression for the one-loop effective Lagrangian
$\cL^{{\rm eff}} (B, \mu, T)$ in  the finite temperature and density QED
is well-known \cite{PRShabad76,Cabo81,ElmforsPS93}.
Nevertheless, at specific
values $\mu, T$ and $B$ this expression may be calculated
only numerically: one
can get an analytical answer for plasma- (i.e. $\mu$- and $T$-dependent)
part of the effective action $\tilde\cL^{\rm eff}(B,\mu,T)$ in the $\mu$
or $B
\rightarrow \infty$ limits \cite{ElmforsPS93} or the  $T \rightarrow 0$
limit
 \cite{PerZeit}, while calculation of $\tilde\cL^{{\rm eff}} (B, \mu, T)$ at
intermediate values of parameters is difficult.  In our previous paper
\cite{Zeitlin3} we have shown that it is possible to obtain
a low-temperature
expansion of the one-loop effective Lagrangian keeping $\mu$ and $B$ finite.
Low-temperature corrections (as well as $\mu$-dependent part at $T=0$) may be
expressed in terms of elementary functions as a finite sum over excited
Landau levels.

The situation with the one-loop polarization operator
$\Pi_{\mu\nu}(p)$  is similar
to that with the effective action. Indeed, a general expression for the
one-loop polarization operator at $B, \mu, T \ne 0$ was obtained in
\refs{PRShabad79,PRShabad82}.  It was presented as a decomposition over six
covariant transversal  tensor structures, but the scalar coefficients had
a form  much more complicated than $\tilde\cL^{\rm eff}(B,\mu,T)$ itself.
On the other hand,
for some applications one needs only  few components of the polarization
operator in
the static limit ($p_0=0, {\bf p} \rightarrow 0$), not the whole
answer. Nevertheless, even the calculation of the static limit
from the general expression possess serious problems. At the same time,
some components of the
polarization operator may be obtained from the fermion density as
derivatives with respect to $\mu$ and $B$ and  as long as the fermion
density may be presented in the simple form, one may obtain a simple
expression  for the above components, too.
Such calculations were performed in \refc{Zeitlin3} to get the
components $\Pi_{00},
{}~\Pi_{01}=\Pi_{10}^*$ and $\Pi_{02}=\Pi_{20}^*$ at a low (and zero)
temperature in the static limit, which have been obtained  as
finite sums of the elementary functions (these calculation were
confirmed in \refc{DGrasso} where the above components had been calculated at
$T=0$ as corresponding one-loop integrals at zero external four-momentum).

In this paper we are continuing the study of the finite density QED
with a uniform
magnetic field at low temperature started  in \refc{Zeitlin3}. We shall
demonstrate that in a finite temperature and density QED in a strong uniform
magnetic field a low-temperature expansion may be performed for a wider range
of parameters $\mu$ and $B$. We
shall pay special attention to show how the finite temperature improves
analytical properties of the fermion density and related quantities, which are
piecewise continuous at $T=0$ \cite{PerZeit,Zeitlin3}.  We shall also
demonstrate that starting from the effective Lagrangian and using the general
properties of the polarization operator one can obtain a static screening in
the finite temperature and density QED with a strong magnetic field. As an
example we shall calculate the Debye screening for some simple charge
configurations to demonstrate that magnetic field splits Debye radius
in a  transversal and a longitudinal parts.

The paper is organized as follows: in Section 2 a low-temperature expansion of
the one-loop effective Lagrangian is obtained for the  partially filled
{\it highest} excited Landau level. In Section~3  the
components of the polarization operator in the static limit are obtained. In
Section 4 the approach used for a partially filled highest Landau level is
extended for the Landau level edge crossing the Fermi surface. We
calculate the Hall
conductivity and show how finite temperature smoothens down the inverse
square-root singularities. In Section 5 the Debye screening in the QED
plasma with external magnetic field is studied.

 \section{Low-temperature expansion in the QED with a magnetic field}

We shall  consider a finite density  QED with a uniform
magnetic field.  At nonzero chemical potential
the corresponding Lagrangian is\footnote{In this paper we shall keep the same
notations as in \refs{PerZeit,Zeitlin3}.
We take the external magnetic field to be parallel to the
$z$--axis, $F_{12}=-F_{21}=B$; $\mu$, $eB >0$.}:

\be
	\cL = -\frac14 F_{\mu\nu}F^{\mu\nu} +\bar{\psi}(i {\partial
         \kern-0.5em/} -e {A\kern-0.5em/}  - \gamma_0 \mu -m)\psi~~~.
\label{tree}
\ee

The one-loop effective Lagrangian at
$T,~\mu,~B \ne 0$, ~$\cL^{{\rm eff}} (B, \mu, T)$
may be written as follows \cite{PRShabad76,Cabo81,ElmforsPS93}:

        \be
        \cL^{{\rm eff}} (B, \mu, T) =   \cL^{\rm eff}(B)+
        \tilde\cL^{\rm eff}(B,\mu,T)~~~,
        \ee

\bs
\noi
where $\tilde\cL^{\rm eff}(B,\mu,T)$,

        \bea
        \tilde\cL^{\rm eff}(B,\mu,T)=
        \frac1{\beta}\frac{eB}{(2\pi)^2} \sum_{k=0}^\infty
        b_k \int_{-\infty}^\infty d\!p_\parallel \left\{
        \ln[ 1+e^{-\beta(\varepsilon_k(p_\parallel)-\mu)}] +
        \ln[ 1+e^{-\beta(\varepsilon_k(p_\parallel)+\mu)}] \right\}
        \label{lagr1}
\eea

\bs
\noi
is  the contribution due to the finite temperature and density
($\beta= 1/T$, ~$p_\parallel$ is the modulus of the momentum parallel to the
magnetic field, $\ve_k(p_\parallel) = \sqrt{m^2 + 2eBk +
p_\parallel^2}$, ~$b_k\equiv 2- \delta_{n,0}~$) and $\cL^{\rm eff}(B)$,

        \be
        \cL^{{\rm eff}} (B)=-\frac{1}{8\pi^2} \int_0^\infty \frac{ds}{s^3}
        \left[ eBs \coth(eBs)-1-\frac13 (eBs)^2 \right]
        \exp(-m^2s)
        \ee

\bs
\noi
is the Euler-Heisenberg effective  Lagrangian in  the purely magnetic
case \cite{ItzyksonZ}.

Integrating  \Eq{lagr1} by parts one has:

        \be
        \tilde\cL^{\rm eff}(B,\mu,T)=
        \frac{eB}{(2\pi)^2} \sum_{n=0}^\infty
        b_n \int_{-\infty}^\infty d\! \pp ~\frac{\pp^2}{\ve_n(\pp)}
        ~(f_+(T) + f_-(T))~~~,
        \label{lagr2}
        \ee

\bs
\noi
$f_\pm(T)$ denotes the Fermi distribution,

        \be
        f_\pm(T) = \frac1{1+e^{\beta(\ve\mp \mu)}} \quad .
        \nonumber
        \ee

Using representation (\ref{lagr2}) one can easily obtain a zero temperature
limit:  at $T \rightarrow 0$ the Fermi distribution approaches the
step-function, $\lim_{T \rightarrow 0} f_\pm = \theta(\pm\mu - \ve)$ and
\Eq{lagr2} reads~\cite{PerZeit}:
        \bea
        \lefteqn{\tilde\cL^{\rm eff}(T = 0,B,\mu)=}\label{zeroT}\\
        &&  \frac{eB}{(2\pi)^2}
        \sum_{n=0}^{ \left[   \frac{\mu^2 - m^2}{2eB}   \right]}
        b_n
        \left\{ \mu
        \sqrt{\mu^2-m^2-2eBn}
        - (m^2+2eBn)
        \ln     \left(
        \frac{ \mu + \sqrt{\mu^2-m^2-2eBn}} {\sqrt{m^2+2eBn}}
                                              \right) \right\},\nonumber
        \eea

\bs
\noi
where $[ \dots ]$ denotes the integral part. To evaluate a low temperature
expansion we make change a of variables and integrate \Eq{lagr2} by
parts again:

        \bea
        \lefteqn{\tilde\cL^{\rm eff}(T,B,\mu)=}\nonumber\\
        &&\frac{eB}{(2\pi)^2}
        \sum_{n=0}^{\infty}
        b_n
        \Bigg\{
        \Big[ \ve \sqrt{\ve^2 - m^2 -2eBn} -\Bigg.\Big.\nonumber\\
        &&\left.\Big.
         (m^2 + 2eBn) \log (\ve + \sqrt{\ve^2 - m^2 -2eBn})  \Big]
        (f_+ + f_-) \right|_{\sqrt{m^2+2eBn}}^\infty -
        \label{lowT1}\\
        && \int_{\sqrt{m^2+2eBn}}^\infty d\!\ve
        \left[ \ve \sqrt{\ve^2 - m^2 -2eBn} ~- \right. \nonumber\\
        &&\left.\left.(m^2 + 2eBn) \log (\ve +  \sqrt{\ve^2 - m^2 -2eBn})
\right]
        \left(\frac{\d f_+}{\d \ve} + \frac{\d f_-}{\d \ve}
                                                \right)\right\}.\nonumber
        \eea

At low temperature the contributions from $\frac{\d f_-}{\d \ve}$,
{}~$f_-$ and from
the upper limit in the first term are exponentially small ($\mu >0$). Making
change of variables and rewriting    the derivative of the Fermi
distribution as
$\frac{\d f_+}{\d \ve} = -\frac1{4T} \cosh^{-2}(\frac{\ve - \mu}{2T})$
we may  present \Eq{lowT1} as follows:
        \bea
        \lefteqn{\tilde\cL^{\rm eff}(T,B,\mu)=}\nonumber\\
        &&\frac{eB}{(2\pi)^2}
        \sum_{n=0}^{\infty}
        b_n
        \Bigg\{
        -(m^2 + 2eBn)\log \sqrt{m^2+2eBn} \theta(\mu - \sqrt{m^2+2eBn})
        +\Bigg.
        \label{lowT2}\\
        &&  \int_{\sqrt{m^2+2eBn} - \mu}^\infty
        d\!q \Bigg[ (\mu +q)  \sqrt{(\mu + q)^2 - m^2 -2eBn} ~- \Bigg.
        \nonumber\\
        &&\Bigg.\Bigg.
        (m^2 + 2eBn) \log \Big((\mu + q) +  \sqrt{(\mu + q)^2 - m^2 -2eBn}\Big)
                                                \Bigg]
        \cosh^{-2}\Big(\frac{q}{2T}\Big)\Bigg\}.\nonumber
        \eea

At low temperature the derivative of the Fermi distribution approaches the
$\delta$-function and in the $T \rightarrow 0$ limit we arrive at \Eq{zeroT}.
To get low-temperature  corrections one may extend lower boundary of the
integration in \Eq{lowT1} to $- \infty$ (the function $\frac1{4T}
\cosh^{-2}(\frac{q}{2T})$ decreases sharply as one moves off the  point $q=0$).
Then one should expand the expression in the brackets in a Taylor
series at $ q=0$ to obtain in the leading order the following
low-temperature correction to the zero-temperature Lagrangian \cite{Zeitlin3}:

        \be
        \Delta\tilde\cL^{\rm eff}(T,B,\mu)= \frac{eBT^2}{6}
        \sum_{n=0}^{ \left[   \frac{\mu^2 - m^2}{2eB}   \right]}
        b_n \frac{\mu}{(\mu^2 - m^2 - 2eBn)^{1/2}}
        + O(T^4)~~~.
        \ee

\noi
Hence,   only those Landau levels which have the edge
laying below the Fermi surface give rise to the effective
Lagrangian at a low temperature.

The above expansion is valid as long as

        \be
        \frac{T}{\mu - \sqrt{m^2 + 2eBn}} \ll 1~~~,
        \label{limit}
        \ee

\noi
which means that the distance from the edge of any Landau level $\ve_k(\pp) =
\sqrt{m^2 + 2eBk}$ to the Fermi surface $\mu$ is much greater than
the temperature (this means also that the Landau level with $n =
\Big[\frac{\mu^2-m^2}{2eB}\Big]$ ({\it highest} excited) should be partially
filled, i.e. its edge cannot coincide with the Fermi surface).

\bigskip
\section{Fermion density and polarization operator}

Having the expressions for the effective Lagrangian
we may move forward to calculate the fermion density
$\rho = \dfrac{\d \cL^{\rm eff}}{\d \mu}$, the magnetization
$M=\dfrac{\partial \cL^{\rm eff}}{\partial B}$,
the Hall conductivity and some components of the polarization
operator in the static
limit ~$p_0=0, ~{\bf p} \rightarrow 0$ \cite{Zeitlin3}.

Using Eqs. (\ref{zeroT}), (\ref{lowT2}) and definitions of the fermion density
$\rho = \dfrac{\d \cL^{\rm eff}}{\d \mu}$ one has:

        \bea
        \lefteqn{\rho(B,\mu,T) =}\label{rho}\\ &&
        \frac{eB}{2\pi^2} \sum_{n=0}^{\left[  \frac{\mu^2 - m^2}{2eB} \right]}
b_n
        \sqrt{\mu^2-m^2-2eBn} \left\{ 1 - \frac{T^2\pi^2}6 \frac{m^2 +
        2eBn}{(\mu^2-m^2-2eBn)^2}      \right\} + O(T^4)~.  \nonumber
        \eea

Then, we may calculate five components of the polarization operator in the
static limit $p_0=0,~{\bf p}\rightarrow 0$. As it was shown in
\refc{Fradkin}, the $\Pi_{00}$--component of the polarization operator may be
written in the static limit as a derivative of the fermion density with respect
to the chemical potential, ~$\Pi_{00}(p_0=0,~{\bf p}\rightarrow 0) =
e^2 \dfrac{\d \rho}{\d \mu}$,

        \bea
        \lefteqn{\Pi_{00}(p_0=0,~{\bf p}\rightarrow 0)=}\label{pi00}\\
        &&e^2
        \frac{eB\mu}{2\pi^2}
        \sum_{n=0}^{\left[  \frac{\mu^2 - m^2}{2eB} \right]}
        b_n
        \left\{
        (\mu^2-m^2-2eBn)^{- 1/2}
        + T^2\pi^2
        \frac{m^2 + 2eBn}{(\mu^2-m^2-2eBn)^{5/2}}         \right\}~~~.
        \nonumber
        \eea

\bs
At $B=0$ ~$\Pi_{00}(p_0=0,{\bf p}\rightarrow 0)$ defines the Debye screening
radius, $r_D^{-2} =\Pi_{00}(p_0=0,{\bf p}\rightarrow \nolinebreak 0)$
\cite{Fradkin} but this relation does not hold for $\mu, B\ne 0$ as the tensor
structure of the polarization operator is more complicated now (modification of
the Debye screening will be discussed in Section 5. See also an instructive
QED$_{2+1}$ example \cite{Pisarsky}).

The components $\Pi_{01}=\Pi_{10}^*$ and $\Pi_{02}=\Pi_{20}^*$  in the static
limit may be expressed via  derivatives of the fermion density with respect
to magnetic field:

        \be
        \Pi_{0j}(p \rightarrow 0) = \i e\/ \varepsilon_{ij}p_i\frac{\d
        \rho}{\d B} \quad i,j = 1,2~~~,
        \ee

\noi
which follows from the definition of the polarization operator,

        \be
        \Pi_{\mu\nu}(x,x') =
        \i~ \frac{\delta <j_\mu (x)>}{\delta A_\nu (x')}~~~.
        \ee

Analyzing the general structure and symmetry properties of the polarization
operator  we may readily define one of the scalar coefficients of the
polarization
operator. The polarization tensor in QED$_{3+1}$ with a uniform magnetic field
at $T,\mu \ne 0$ may be decomposed over six tensor structures
\cite{PRShabad79} (we are using a slightly modified
with respect to \refs{PRShabad79,Zeitlin3} expression):

        \bea
        \lefteqn{\Pi_{\mu\nu}(p|T,\mu,B)=}\nonumber\\
        &&\nonumber\\
        &&\left(
        g_{\mu\nu} - \frac{p_\mu p_\nu}{p^2}               \right){\cal A}~
        +\left(
        \frac{p_\mu p_\nu}{p^2} -
        \frac{p_\mu u_\nu + u_\mu p_\nu}{(pu)} +
        \frac{u_\mu u_\nu}{(pu)^2} p^2                          \right)
        {\cal B} + \nonumber\\
        &&\nonumber\\
        &&
        F_{\mu\alpha}F^{\alpha\beta}p_\beta
        F_{\nu\phi}F^{\phi\rho}p_\rho
        ~{\cal C}
        + F_{\mu\lambda}p^\lambda F_{\nu\phi}p^\phi {\cal D} +\label{pimunu}\\
        &&\nonumber\\
        &&  \i \left(
        p_\mu F_{\nu\lambda}p^\lambda  -
        p_\nu F_{\mu\lambda}p^\lambda  +
        p^2F_{\mu\nu}                   \right) {\cal E}~
        + \i \left(
        u_\mu F_{\nu\lambda}p^\lambda  -
        u_\nu F_{\mu\lambda}p^\lambda  +
        (pu)F_{\mu\nu}                   \right) {\cal F}~~~.\nonumber
        \eea

\bs
The scalars ${\cal A, B, C, D, E}$ and ${\cal F}$ are the functions
of $p_0^2, ~p_3^2=p_\parallel^2, ~p_1^2 + p_2^2=p_\perp^2$ and $B$ ($u^\mu$ is
the $4$-velocity of the medium \cite{Fradkin}, $u^\mu = (1,0,0,0)$) , therefore
only the last tensor structure may contribute to $\Pi_{0j}$
and  we may define the coefficient ${\cal F}$ in the static limit
(${\cal F} \ne 0$ at $B,\mu \ne 0$ only \cite{Shabad88}):

        \be
        {\cal F}(p_0=0,~{\bf p}\rightarrow 0) = \frac{e}{B}
        \frac{\d \rho}{\d B}~~~.
        \ee

It is easy to see that the components $\Pi_{0j}$  describe a
conductivity in the plane orthogonal to the magnetic field  which is
Hall-like \cite{Zeitlin3,Zeitlin2}:

        \be
        \sigma_{ij} =
        \left. \frac{\d j_i}{\d E_j} \right|_{E \rightarrow 0}=
        \i \left. \frac{\d \Pi_{0i}(p)}{\d p_j}
                \right|_{p \rightarrow 0}
        = e \varepsilon_{ij} \frac{\d \rho}{\d B}
        \quad i,j = 1,2 ~~~.
        \label{sigma}
        \ee

\bs
Substituting into \Eq{sigma} the expression for the fermion density,
\Eq{rho} one has

        \bea
        \lefteqn{\Pi_{0j}(p_0, {\bf p} \rightarrow 0) =
        \frac{\i \varepsilon_{ij}p_i}{2\pi^2}
        \sum_{n=0}^{\left[  \frac{\mu^2 - m^2}{2eB} \right]}b_n
        \times}\nonumber\\
        && \left\{
        \frac{\mu^2-m^2-3eBn}{(\mu^2-m^2-2eBn)^{1/2}} -
        \frac{T^2\pi^2}3
        \frac{(\mu^2-m^2-2eBn)(m^2 + 2eBn) + 3eBn\mu^2}{(\mu^2-m^2-2eBn)^{5/2}}
                                        \right\}~~~.
        \label{hall}
        \eea

It follows from the above expression that the Hall conductivity in the
QED$_{3+1}$ is an oscillating function of the chemical potential and the
magnetic field and in the $T \rightarrow 0$ limit it
has an inverse square-root singularity (we would like to remind that
polarization operator in QED$_{3+1}$ with a uniform magnetic field has just the
same kind of singularities \cite{PRShabad82}).  These oscillations are close to
``giant oscillations'', well-known in condensed matter physics \cite{Abrikosov}
and resonant effects in QED \cite{SUsov,Shabad72,Shabad75} and semiconductors
\cite{KShabad,ZShabad}.
In the next Section we shall calculate
finite-temperature corrections at the points $\mu \rightarrow \sqrt{m^2 +
2eBk}$ to show how the temperature cures the singularities.

\section{Low-temperature expansion at $\mu = (m^2 + 2eBk)^{1/2}$}

Now we shall show how a low-temperature expansion at
$\mu \rightarrow \sqrt{m^2 + 2eBn}$  may be derived.
At $T=0$ the functions, calculated in the previous Section are not smooth and
components of the polarization operator even do not possess a continuous
limit. As an example we shall consider the Hall conductivity,
$e\frac{\d \rho}{\d B}$.

Differentiating effective Lagrangian \Eq{lagr1} with respect to $\mu$ to obtain
the fermion density at arbitrary $T$ and taking derivative with
respect to $B$
(we assume below $T$ to be small, thus  $f_-$-dependent part may be omitted)
one has:

        \be
        \sigma= \sigma_{(1)} + \sigma_{(2)}=
        \frac{e}{(2\pi)^2} \sum_{k=0}^\infty b_k
        \int d\!p_\parallel f_+
        \frac{eB}{(2\pi)^2} +
        \sum_{k=0}^\infty b_k \int d\!p_\parallel
        \frac{\d f_+}{\d \ve_k}\frac{\d \ve_k}{\d B}~~~.
        \label{hallT}
        \ee

Let us consider $\sigma_{(2)}$  first. Making change of variables one gets:

        \be
        \sigma_{(2)} =
        \frac{eB}{\pi^2}
        \sum_{k=1}^\infty b_k
        \int_{\sqrt{m^2+2eBk}-\mu}^\infty d\!z
        \frac1{\sqrt{(z+\mu)^2 - m^2 -2eBk}}
        \left(- \frac1{4T} \right) \cosh^{-2} \left( \frac{z}{2T} \right)~~~.
        \label{sigma2.2}
        \ee

Assuming $\mu \rightarrow \sqrt{m^2 + 2eBk_0}$ and extracting the
$k_0$-term
from the sum  (\ref{sigma2.2}) (the remaining contribution to $\sigma_{(2)} $
may be derived along the same lines as in Section 2) one has:

        \be
        \lim \phantom{.}_{\mu \rightarrow ~\sqrt{m^2 +  2eBk_0}}
        \sigma_{(2)}^{k_0} = - \frac{eB}{4\pi^2 T} ek_0
        \int_0^\infty d\!z \frac{z^{-1/2}}{\sqrt{z+2\mu}} \cosh^{-2} \left(
        \frac{z}{2T} \right)
        \label{sigma2.3} \ee

Expanding the inverse square root in \Eq{sigma2.3} at $z=0$ we finally
obtain:

        \be
        \sigma_{(2)}^{k_0} = - \frac{e(2^{3/2}-1)\zeta({1\over 2})}
        {8\pi^{5/2}} \frac{\mu^2-m^2}{\sqrt{2T\mu}}~~~.
        \ee

Making a similar calculation for $\sigma_{(1)}$  one has
$\sigma_{(1)} \sim \sqrt{T}$.
Therefore, at a finite temperature the inverse square-root
singularity in the expression for the Hall conductivity (as well as for
above-mentioned components of the polarization operator) is
smoothened down.  The leading
contribution in the low-temperature limit is finite and proportional to
$T^{-1/2}$. Making the similar expansions for the fermion density and
magnetization  one may see
that the finite-temperature correction levels corresponding
zero-temperature expressions (cf.  \cite{ElmforsPS93,PerZeit}).

\section{Debye screening via effective action}

The most interesting application of the above analysis is the possibility to
study the static screening properties of the magnetized nondissipative
relativistic
electron-positron gas.  To study the screening the gauge field Green function
${\cal D}_{\mu\nu}$ with appropriate quantum corrections  is required,

        \be
        {\cal D}_{\mu\nu} =
         \left(
        D_{\mu\nu}^{-1} - \Pi_{\mu\nu}          \right)^{-1}~~~,
        \ee

\noi
$D_{\mu\nu}$ is
the tree-level propagator.

Formally, there are no obstacles to obtain an expression for ${\cal
D}_{\mu\nu}$ by
inverting the corresponding matrix, but with $\Pi_{\mu\nu}$ as complicated
as in
\Eq{pimunu} it requires unwieldy algebra. Another possibility to
calculate ${\cal D}_{\mu\nu}$ is to solve the eigenvalue
equation, $\Pi_{\mu\nu}(p)b^\nu = \kappa
b_\mu$ and to rewrite $\Pi_{\mu\nu}(p)$ in the diagonal form
\cite{BShabad},

        \be
        \Pi_{\mu\nu}(p) = \sum_{i=1}^3 \kappa_i
        \frac{b_\mu^{(i)}b_\nu^{(i)*}}{b_\alpha^{(i)} b^{\alpha(i)*}} \quad .
        \ee

\noi
Then, one readily obtains

        \be
        {\cal D}_{\mu\nu}(p) = \sum_{i=1}^3
        \frac1{p^2 -\kappa_i} \frac{b_\mu^{(i)}
        b_\nu^{(i)*}} {b_\alpha^{(i)} b^{\alpha(i)*}} \quad ,
        \ee

\noi
but this representation either does not  simplify the calculation of a
specific component of ${\cal D}_{\mu\nu}$.

To investigate the Debye screening one needs ${\cal D}_{00}(p_0=0,{\bf
p})$. For this particular case we have made the relevant calculations
in the Feynman gauge keeping all structures contained in the
expression (\ref{pimunu}).  We have no possibility to present this cumbersome
result and  show the static limit of the expression only.
The coefficients ${\cal A, C, D}$ and ${\cal E}$ happen to be
subleading in ${\bf p}^2$ (see also \refc{DGrasso}) and may lead to some finite
corrections (of order $e^2$) which do not affect the qualitative picture.
The
final expression may be written as follows ($\tilde{\cal F} = B {\cal F}$):

        \be
        {\cal D}_{00} =
        -\frac{{\bf p^2}}{(\Pi_{00}+{\bf p^2}){\bf p^2}+
        \tilde{\cal F}^2 p^2_\perp}
        \ee

The Coulomb potential,

        \be
        A_0({\bf x}) = \int d\!{\bf p} e^{-\i {\bf px}}
        {\cal D}_{00}(p_0=0, {\bf p}) j_0({\bf px})
        \label{coulomb}
        \ee

\noi
in the case of a point-like test charge may be calculated only
numerically so
we shall study examples of simpler charge configurations (i.e. these
reducing
one of momentum integrations in \Eq{coulomb}).  For instance, let us consider
the field of an  infinitely long charged thin rod parallel to $z$-axis. In
this case

        \be
        A_0(r) = \rho_c \int_0^\infty p_\perp d\! p_\perp
        \int_0^{2\pi}  d\!\phi e^{-\i p_\perp r \cos \phi}
        \frac1{\Pi_{00}+\tilde{\cal F}^2 + p^2_\perp}
        = \rho_c {\rm K}_0(r\sqrt{\Pi_{00}+\tilde{\cal F}^2})~~~,
        \ee

\noi
K$_0$ is the modified Bessel function,
$\lim_{r\rightarrow\infty}{\rm K}_0(r)\sim
\sqrt{{\pi \over 2r}} e^{-r}$ (cf. screening in QED$_{2+1}$, \refc{PRao}).

Therefore, in the presence of a strong background magnetic field the Debye
screening is described by at least two scalar functions and the
suggestion that
$\Pi_{00}$ only is responsible for screening at $B, \mu \ne 0$  made in
\refc{DGrasso} is incorrect.

Then, let us consider potentials of charged planes parallel and orthogonal to
magnetic field:

        \be
        A_0(x_1) = \rho_c \int d\! p_1
        \frac{e^{-\i p_1 x_1}}{\Pi_{00}+\tilde{\cal F}^2 + p^2_1} =
         \frac{\pi\rho_c}{\sqrt{\Pi_{00}+\tilde{\cal F}^2}}
        e^{- x_1 \sqrt{\Pi_{00}+\tilde{\cal F}^2}}~~~,
        \ee

        \be
        A_0(z) =\rho_c  \int d\! p_\parallel
        \frac{e^{-\i p_\parallel z}}{\Pi_{00}+p^2_\parallel} =
        \frac{\pi\rho_c }{\sqrt{\Pi_{00}}}
        e^{- z \sqrt{\Pi_{00}}}~~~,
        \ee

\noi
We may see that the Debye radius is splitted in plasma with external
magnetic
field, $r_D^\parallel \ne r_D^\perp$, therefore, screening is anisotropic
(in \refc{HSivak} it was
shown that the
small-distance screening depends on direction at $\mu = 0$ and high $T$).
We would like to stress that at a realistic values of parameters
the  long distance
anisotropy is quantitatively small (in the vicinity of Landau level edge
$\Pi_{00}$ and $\tilde{\cal F}^2$ may be of the same order, but in that
case the
screening radius will be much less than an average distance between the
particles and the Debye approximation fails). In the above-described
examples the
anisotropy in the leading order is due to additional tensor structure only
(cf. resonant effects described in \refs{SUsov,Shabad88,KShabad,ZShabad}).

Supposing that the coefficients ${\cal A, C, D}$ and ${\cal E}$ will
not affect qualitatively the other components of ${\cal D}_{\mu\nu}$
we may write the latter in the static limit as follows (we are using
the Feynman gauge):

         $$
        {\cal D}_{\mu\nu}^{-1} =
        \left(
        \begin{array}{cccc}
        -(\Pi_{00} +{\bf p}^2) & \i \tilde{\cal F}p_2 \quad&
        -\i \tilde{\cal F}p_1 \quad & 0\\
        -\i \tilde{\cal F}p_2 &{\bf p}^2 &0&0\\
        \i \tilde{\cal F}p_1 & 0& {\bf p}^2&0\\
        0&0&0&{\bf p}^2
        \end{array}\right)
        \eqno{(33)}
        $$

It is easy to check that antysimmetric $p$-linear structure in the
polarization operator, unlike Chern-Simons term in QED$_{2+1}$ \cite{DJT}, does
not lead to magnetic screening (we would like to stress that with the
polarization operator \Eq{pimunu} condition $\Pi_{ii}=0$ \cite{GPY} does not
guarantee the absence of the magnetic mass).

\bigskip
We have demonstrated that it is possible to perform a low-temperature expansion
in QED$_{3+1}$ with a uniform magnetic field. In the framework of this
expansion both effective Lagrangian and derivative functions have the same
structure as in the $T=0$ limit, i.e. they are finite sums over excited
Landau
levels. We have calculated the components of the polarization operator
responsible
for the Hall conductivity and Debye screening in a finite fermion density
QED$_{3+1}$
with a uniform magnetic field. We have shown how finite temperature cures
zero-temperature singularities. Static potential of some charge configurations
was calculated and anisotropy of the Debye screening was demonstrated.

\bigskip \section*{Acknowledgement} I am  grateful to
V.~V.~Losyakov, A.~M.~Semikhatov and I.~V.~Tyutin for fruitful discussions and
to Vl.~Zeitlin for critical reading of the manuscript.  This research was made
possible in part by Grant $N^o$ MQM300 from the International Science
Foundation and Government of Russian Federation and RBRF Grant $N^o$
93-02-15541.

\end{document}